\documentclass[twocolumn,aps,prb,superscriptaddress,longbibliography]{revtex4-1}
\usepackage{graphicx,amssymb,amsmath,psfrag,xcolor,bm,xfrac}
\usepackage[colorlinks=true,citecolor=blue,linkcolor=blue,urlcolor=blue]{hyperref}

\usepackage[normalem]{ulem} 
\begin{document}
	\title{Strong quantum interactions prevent quasiparticle decay}
	\author{Ruben Verresen}
	\affiliation{Department of Physics, T42, Technische Universit\"at M\"unchen, 85748 Garching, Germany}
	\affiliation{Max-Planck-Institute for the Physics of Complex Systems, 01187 Dresden, Germany}
	\affiliation{Kavli Institute for Theoretical Physics, University of California, Santa Barbara, CA 93106, USA}
	\author{Frank Pollmann}
	\affiliation{Department of Physics, T42, Technische Universit\"at M\"unchen, 85748 Garching, Germany}
	\date{\today}
	\author{Roderich Moessner}
	\affiliation{Max-Planck-Institute for the Physics of Complex Systems, 01187 Dresden, Germany}
	
\begin{abstract}
	    Quantum states of matter---such as solids, magnets and topological phases---typically exhibit collective excitations---phonons, magnons, anyons\cite{Venema16}. These involve the motion of many particles in the system, yet, remarkably, act like a single emergent entity---a \emph{quasiparticle}. Known to be long-lived at the lowest energies, common wisdom says that quasiparticles become unstable when they encounter the inevitable continuum of many-particle excited states at high energies. Whilst correct for weak interactions, we show that this is far from the whole story: \emph{strong} interactions \emph{generically} stabilise quasiparticles by pushing them out of the continuum. This general mechanism is straightforwardly illustrated in an exactly solvable model. Using state-of-the-art numerics, we find it at work also in the spin-$\sfrac{1}{2}$ triangular lattice Heisenberg antiferromagnet (TLHAF) near the isotropic point---this is surprising given the common expectation of magnon decay in this paradigmatic frustrated magnet. Turning to \emph{existing} experimental data, we identify the detailed phenomenology of avoided decay in the TLHAF material Ba$_3$CoSb$_2$O$_9$, and even in liquid helium---one of the earliest instances of quasiparticle decay\cite{Pitaevskii59}. Our work  unifies various phenomena above the universal low-energy regime in a comprehensive description. This broadens our window of understanding of many-body excitations, and provides a new perspective for controlling and stabilising quantum matter in the strongly-interacting regime.
\end{abstract}
	
	\maketitle
	
It is a fundamental insight of quantum mechanics that energy levels repel.
This is commonly illustrated by letting two levels with unperturbed (`bare') energies $\pm E_b$ interact with one another through a coupling $\gamma$, i.e.
\begin{equation}
\hat H = 
\left( \begin{array}{cc}
E_b & \gamma \\
\gamma & - E_b
\end{array} \right). \label{eq:twobytwo}
\end{equation}	
The resulting energies of $\hat H$ are $\pm \sqrt{E_b^2 + \gamma^2}$. Hence, repulsion leads to a minimal separation of the levels of $2|\gamma|$, no matter how small the initial separation $2|E_b|$.

A natural question is whether this extends to the case of a discrete level coupled to a continuum of states.
The question might seem moot, since the common expectation is that a bare level inside a continuum will be \emph{dissolved} by interactions.
At best, it will become a finite-lifetime resonance. At worst, no hint of it remains. 

If the bare level represents a quasiparticle,
its broadening and  disappearance in the many-particle continuum is known as quasiparticle decay. In the case of non-topological\footnote{i.e. an ordered magnet or a non-topological paramagnet} quantum magnets---where quasiparticles go under the name of magnons, or spin waves---the expectation of magnon decay has, surprisingly only recently, been borne out in inelastic neutron scattering experiments\cite{Stone06,Masuda06,Oh13,Robinson14,Hong17}, see below.

Here, we show that for strong interactions this expectation of quasiparticle decay is wrong.

Rather, 
with increasing  interaction strength, an infinitely long-lived 
state re-emerges out of the
continuum of states. This happens via a 
simple generalisation of the familiar level repulsion, Eq.~\eqref{eq:twobytwo}, for a bare state $|\psi\rangle$ with bare energy $E_b$ coupled to a \emph{continuum} of states $|\varphi_\alpha\rangle$ with bare energies $E_\alpha$ above a threshold energy $E_{\mathrm{th}}$. 
Physically, this model represents states with a fixed value of \emph{total} momentum---the continuous index $\alpha$ corresponds to the \emph{relative} momentum of two-particle states.

Concretely, for large enough coupling $|\gamma|$, there is a single discrete state $|\psi^*\rangle$ with an energy \emph{below} the continuum, $E^*<E_{\mathrm{th}}$ (see Methods). Moreover, the contribution of the unperturbed state $|\psi\rangle$  to this final discrete state, denoted by the weight $Z = |\langle \psi| \psi^*\rangle|^2$, can be large---for a  continuum  occupying a finite range of energy,  the weight approaches $Z \to \sfrac{1}{2}$ for large  $|\gamma|$.

This is experimentally important: a vanishing $Z$ implies that the state $|\psi^*\rangle$ bears little relationship to the original quasiparticle. However, a \emph{large} $Z$ ensures that any experimental set-up---e.g.~neutron scattering---for detecting the original quasiparticle $|\psi\rangle$ also detects $|\psi^*\rangle$.
Hence, while existence of $|\psi^* \rangle$ and \emph{finiteness} of $Z$ for this simple model have been pointed out before \cite{Gaveau95},
its phenomenology and in particular its relevance to quasiparticles in strongly-interacting quantum systems seem to have been underappreciated.

\begin{figure}
		\includegraphics[scale=1]{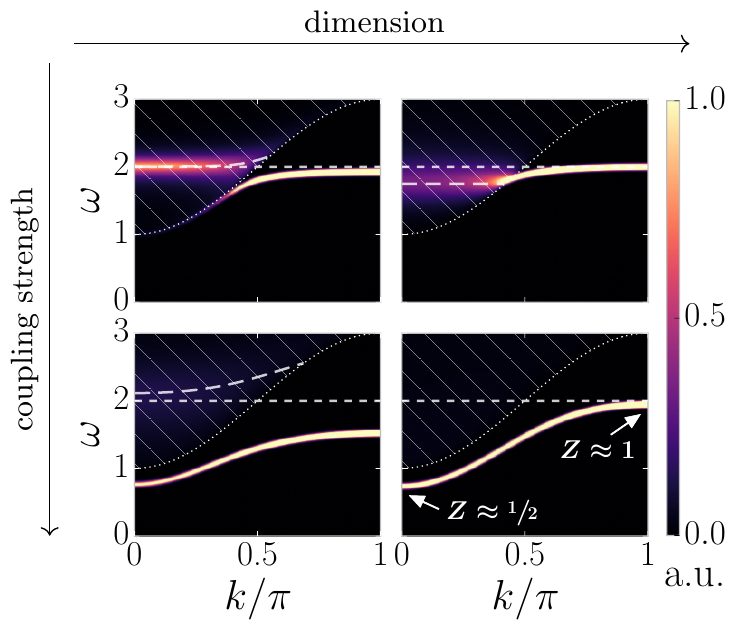}
		\caption{\textbf{Avoided quasiparticle decay in a solvable model.} The bare level $|\psi_k\rangle$ (short-dashed line) is coupled to a continuum (shaded). Left column is representative of gapped spectra in dimensions $D=1,2$; the level cannot enter the continuum, but weight is transferred into a decaying mode (long-dashed line). Right column is for $D\geq 3$. For strong interactions, the outcome is independent of dimension: a renormalised quasiparticle $|\psi_k^*\rangle$ is pushed out, whose weight $Z_k = |\langle \psi_k | \psi_k^* \rangle|^2$ approaches  $\sfrac{1}{2}$. \label{fig:analytical}}
\end{figure}
 
Fig.~\ref{fig:analytical} illustrates what inelastic neutron scattering would measure for a system described by this solvable model (see Methods). It shows the weight of the bare state $|\psi\rangle$ on the true eigenstates. The initially flat bare level (dashed line) is coupled to a continuum (shaded region). For weak interactions, the physics depends on whether the bare energy level encounters a large or small number of states upon entering the continuum. This is encoded in the density of states (DOS), $\nu(E)$. In this example we treat the case of the two-particle continuum of particles with a parabolic dispersion (although any dispersion can be accommodated); its onset satisfies $\nu(E_{th} + \delta E) \sim (\delta E)^{\sfrac{D}{2}-1}$ in $D$ spatial dimensions \cite{Suppl}.

In low dimensions ($D =1,2$), where this DOS  has a discontinuous onset, the infinitely-long lived state $|\psi^*\rangle$ exists for \emph{any} nonzero coupling strength $\gamma$ \cite{Gaveau95}. This might seem surprising as it implies that the $\gamma \to 0$ limit is singular; this is
resolved by the weight $Z$ vanishing as $|\psi^*\rangle$ approaches the continuum (first panel of Fig.~\ref{fig:analytical}).
Most weight is transferred into a decaying mode, and detecting the \emph{residual} quasiparticle requires very high resolution measurements, see below. 
In higher dimensions ($D\geq3$), the quasiparticle enters the continuum more straightforwardly---at least for weak interactions.

Increasing the coupling strength, the quasiparticle re-emerges for any $D$ (Fig.~\ref{fig:analytical} bottom), accompanied by the  weight $Z\rightarrow\sfrac{1}{2}$, in agreement with our general claim.

\begin{figure}
	\includegraphics[scale=1]{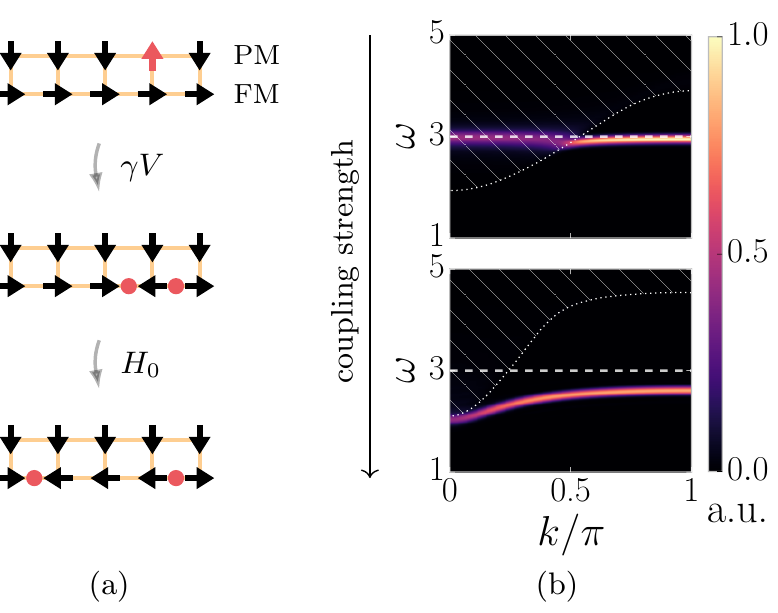}
	\caption{\textbf{Avoided decay in an Ising ladder.} (a) A paramagnet (PM) with magnon-like excitation (red arrow); an Ising ferromagnet (FM) where the quasiparticles are domain walls. By coupling the two chains, a magnon can decay into two domain walls (red dots). (b) The dynamic structure factor; the dashed line is the bare magnon dispersion and the shaded region denotes the continuum of two domain walls. At low coupling strength, the magnon decays. For strong interactions, the magnon is pushed below the continuum.  \label{fig:ladder}}
\end{figure} 
 
How widely applicable is this mechanism of avoided quasiparticle decay? 
Note that the fact we assumed $\gamma$ to be independent of $\alpha$ is not important, since in the full solution, $\gamma^2$ and the DOS always appear together. For example, in a system with $SO(3)$ spin-symmetry, the coupling constant vanishes near the threshold as $\gamma(E_{th} + \delta E) \sim \sqrt{\delta E}$ \cite{Zhitomirsky06}. This leads to a different power of the onset of $\gamma^2 \nu(E)$, which amounts to a simple shift of the effective dimensionality $D \to D+2$, preserving the phenomenology.
Similarly, one could effectively include direct interactions \emph{within} the continuum by using a renormalised DOS.

There are, however, two essential implicit assumptions. Firstly, that there is space below the continuum to be repelled into. This is not applicable to, for example, Fermi liquids, where the continuum starts directly above the ground state energy over an \emph{extended} region in momentum space. 
Second, the model does not actually treat the situation where the continuum is made of the \emph{same} quasiparticles that it repels---making it exactly solvable. This should be a good approximation if the quasiparticle trying to enter the continuum has its momentum $\bm k$ well-separated from those quasiparticles whose momenta $\bm q$ and $\bm k - \bm q$ make up the continuum at that point. As discussed below, this turns out to be the case in the TLHAF.

Before considering the challenging TLHAF, we verify our predictions in a tunable, yet \emph{numerically} tractable, fully many-body quantum system. This consists of two spin-$\frac{1}{2}$ chains: one a perfect paramagnet in a field, $\hat H_0^{(A)} = - 3 \sum_n \hat S^z_{A,n}$, the other an ordered quantum Ising ferromagnet, $\hat H_0^{(B)} = - \sum_n \left( 4J \hat S^x_{A,n} \hat S^x_{A,n+1} + 2g \hat S^z_{A,n} \right)$. 
The ground state of the paramagnet has all spins pointing up; a flipped spin is a dispersionless magnon. The ferromagnet is ordered along the $x$-direction, with freely moving domain wall quasiparticles.

Inter-chain coupling can allow the magnon to decay into a pair of domain walls, illustrated in Fig.~\ref{fig:ladder}(a); for this, consider the interaction $H_\textrm{int} = 4\gamma \sum_n \hat S^x_{A,n} \hat S^z_{B,n}$.
Our numerical data obtained using the dynamical density matrix renormalization group method (DMRG) \cite{White92,Schollwoeck11,Zaletel15}, Fig.~\ref{fig:ladder}(b), confirms the resulting familiar magnon decay for weak interactions. Crucially, as advertised, strong interactions prevent quasiparticle decay: the magnon re-emerges from the continuum unscathed. For precise values of the parameters, see Methods.
 
We now turn to the paradigmatic spin-$\sfrac{1}{2}$ TLHAF, which describes a wide range of frustrated quantum spin materials (see Ref.~\onlinecite{Kojima18} for a recent overview). Its ground state is magnetically ordered, with neighbouring spins forming a 120$^{\circ}$ angle\cite{Huse1988,Bernu1992}. However, mystery enshrouds its magnon excitations due to the uncontrolled nature of the available analytic and numerical methods \cite{Zheng06,Chernyshev09,Mourigal13}. The most venerable of these is perhaps spin wave theory (SWT), an expansion in inverse spin, $1/S$.

We consider the spin-$\sfrac{1}{2}$ TLHAF
\begin{equation}
\hat H = J \sum_{\langle \bm n, \bm m\rangle} \left( (1-\delta) \; \bm{\hat S_n  \cdot \hat S_m} - \delta \; \hat S^{\textrm{loc},z}_{\bm n} \hat S^{\textrm{loc},z}_{\bm m} \right) 
\label{eq:TLHAF}
\end{equation}
where a small easy-axis anisotropy ($\delta=0.05$) slightly gaps out the massless Goldstone modes, making the model more numerically tractable. Here, $\bm{ \hat S_n}^\textrm{\textbf{loc}}$ is the spin in the basis of the rotating (local) frame.

\begin{figure}
\includegraphics[scale=.95,trim=0.1cm 0.1cm 0.22cm 0.1cm,clip]{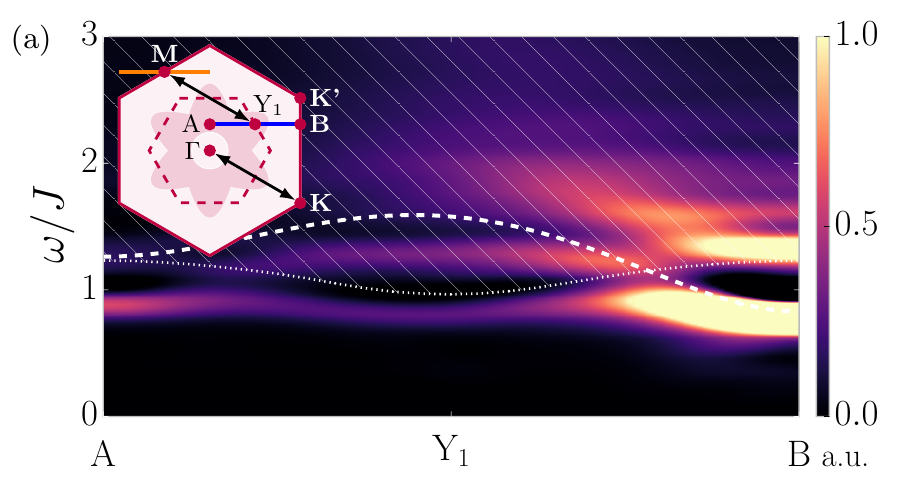}
\includegraphics[scale=.95,trim=0cm 0.1cm 0.1cm 0.05cm,clip]{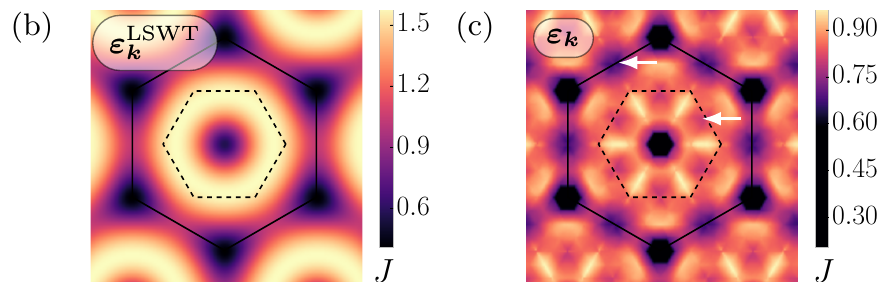}
\caption{\textbf{Avoided decay in the spin-$\bm{\frac{1}{2}}$ TLHAF with $\bm{\delta=0.05}$.} 
(a) Inset: the Brillouin zone; the dashed line delineates the \emph{magnetic} BZ. LSWT predicts magnon decay in the shaded region, dominated by the process $\bm q \to (\bm q - \textrm{K}) + \textrm{K}$. The black arrow illustrates that $\textrm{Y}_1 = \textrm{M} + \textrm{K}$; hence decay is possible if $\varepsilon_{\textrm{Y}_1} > \varepsilon_\textrm{M} + \varepsilon_\textrm{K}$.
Main panel: the out-of-plane dynamical spin structure factor along the blue line in the inset. The dotted line represents the two-magnon states consisting of a magnon along the orange line (inset) and a K-magnon. The dashed line is the magnon dispersion from LSWT. We see avoided decay, where the level-continuum repulsion induces a local minimum near Y$_1$.
(b) The LSWT prediction for the dispersion relation, whereas (c) shows the numerical result. The dispersion is most heavily renormalised where LSWT predicts decay (see inset of (a)).
The local minimum at M induces a local minimum at Y$_1$ (white arrows). \label{fig:TLAFMnum}}
\end{figure}

For this value of $\delta$, SWT predicts magnon decay \cite{Chernyshev09} over a large region of momentum space (shaded region in the inset of Fig.~\ref{fig:TLAFMnum}(a)). A magnon with momentum $\bm k$ is then predicted to decay into two magnons with momenta $\bm q$ and $\bm k - \bm q$, where $\bm q \approx \textrm K$, the corner of the Brillouin zone (BZ). However, small spin and noncollinear order---breaking all symmetries, allowing for many interaction terms---generate strong quantum interactions. Our model thus suggests an alternative to the commonly expected scenario of magnon decay.

A recent advance in numerically simulating the dynamics of two-dimensional quantum systems\cite{Gohlke17,Verresen18} allows to directly test the prediction of magnon decay in Eq.~\eqref{eq:TLHAF}.
Fig.~\ref{fig:TLAFMnum}(a) shows the out-of-plane dynamical spin structure factor along the A--B line (blue line in inset) obtained from dynamical DMRG (see Methods). Since SWT predicts decay into a K-magnon, the dotted line shows the two-magnon energy $\varepsilon_{\bm q} + \varepsilon_{K}$, with $\bm q$ along the orange line in the inset. The dashed curve is the SWT prediction of the magnon in the non-interacting limit $\sfrac{1}{S} \to 0$ (LSWT), traveling deep into the two-magnon continuum. However, the numerically-obtained $S=\sfrac{1}{2}$ dispersion is pushed out completely---a crisp instance of avoided magnon decay.

The dispersion is known to have a local minimum at the midpoint M of the BZ edge. This appears at higher order in SWT and in series expansion methods \cite{Zheng06,Chernyshev09,Mourigal13}, as confirmed in Fig.~\ref{fig:TLAFMnum}(c).
Our \emph{novel} prediction is that the avoided decay must in turn induce a local minimum at the midpoint Y$_1$ of the \emph{magnetic} BZ (MBZ) edge. This is apparent in Fig.~\ref{fig:TLAFMnum}(a,c). More precisely, absence of magnon decay implies the strong constraint $|\varepsilon_{\textrm M} - \varepsilon_{\textrm{Y}_1} | \leq \varepsilon_{\textrm K}$, which we find to be satisfied in our numerics---and in disagreement with SWT. 

\begin{figure}[h]
	\includegraphics[scale=.99,clip,trim=0.1cm 0.2cm 0.1cm 0.15cm]{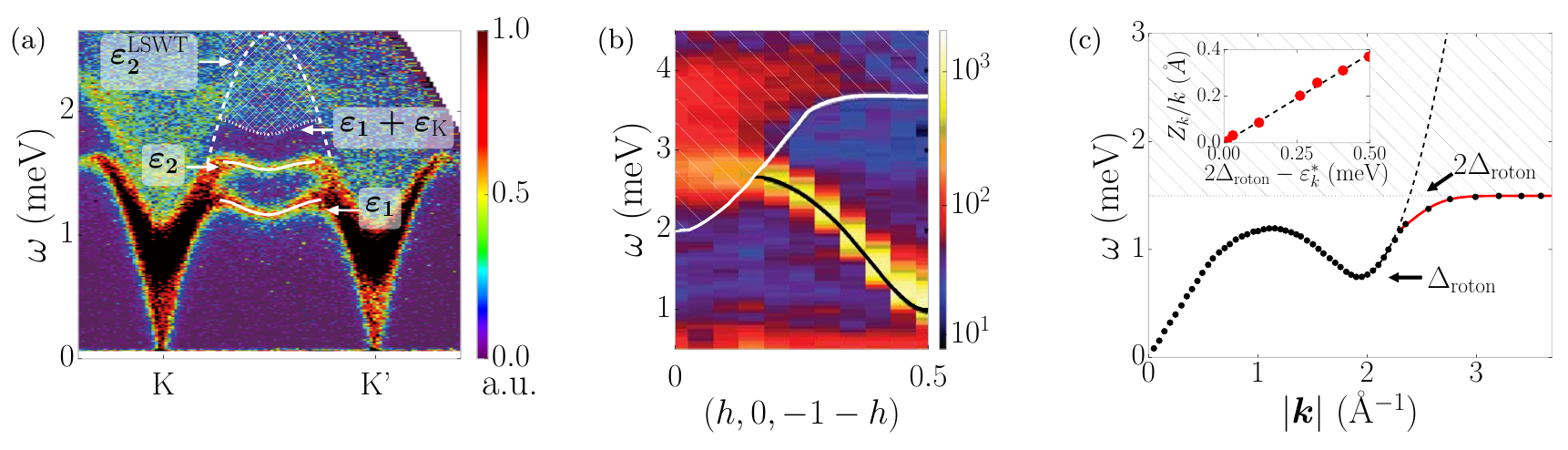}
	\caption{\textbf{Avoided quasiparticle decay, genuine decay, and level-continuum repulsion in experimental data: the TLHAF material Ba$_3$CoSb$_2$O$_9$, PHCC, and superfluid helium.}
	(a) Inelastic neutron scattering data and LSWT comparison for Ba$_3$CoSb$_2$O$_9$ (see Methods for details)\cite{Ito17}. The neutron data picks up all magnon bands related by momentum K; the lower branch ($\varepsilon_1$) goes through M, the higher branch ($\varepsilon_2$) through Y$_1$ (see Fig.~\ref{fig:TLAFMnum} for BZ labeling). Similar to Fig.~\ref{fig:TLAFMnum}, magnon decay is avoided, with the local minimum near M inducing a local minimum near Y$_1$.
	(b) A scenario where the quasiparticle \emph{does} decay: inelastic neutron scattering data\cite{Stone06} for PHCC; white shaded region denotes the two-magnon continuum; black line traces the magnon which decays into the continuum.
	(c) Black dots are the phonon-roton dispersion of superfluid helium extracted from Refs.~\onlinecite{Donnelly1981,Glyde98}; inset shows single-particle weight extracted from Refs.~\onlinecite{Gibbs99,Azuah13} (see Methods). Our solvable model implies that the level approaches the continuum exponentially in the bare level, i.e. $E^*_k \propto - \exp{\left(-b \times k \times E^\textrm{bare}_k\right)}$ (solid red line); here $E^*_k := \varepsilon^*_k-2\Delta_\textrm{roton}$ and $E^\textrm{bare}_k := \varepsilon^\textrm{bare}_k-2\Delta_\textrm{roton}$, with the bare level $\varepsilon^\textrm{bare}_k$ estimated by fitting the roton minimum to a parabola. Moreover, the weight is predicted to go to zero proportional to the level approaching the continuum, i.e. $Z_k \sim a \times k\times |E_k^*|$, confirmed in the inset. In both cases we find that $a_\textrm{fit} \; k_\textrm{roton}$ and $b_\textrm{fit} \;k_\textrm{roton}$ are comparable to the (inverse) bandwidth, in testament to the strong interactions. \label{fig:exp}}
\end{figure}

Intriguingly, this phenomenology \emph{has already been observed} in experiment. The magnetic material Ba$_3$CoSb$_2$O$_9$ is well-described by the TLHAF with a small easy-plane anisotropy, for which Fig~\ref{fig:exp}(a) shows recent inelastic neutron scattering data \cite{Ito17}. Since this is sensitive to the \emph{full} dynamical spin structure factor, it picks up copies of the magnon dispersion translated by K. Fig.~\ref{fig:exp}(a) thus shows \emph{two} bands: the bottom one ($\varepsilon_1$) centered at M, the top one ($\varepsilon_2$) centered at Y$_1$. Neither decay and both exhibit a local minimum, in agreement with the phenomenology of Fig.~\ref{fig:TLAFMnum}. We can thus directly reinterpret apparently unrelated experimental features 
as having a joint origin in avoided quasiparticle decay. 

In contrast, magnon decay \emph{has} experimentally been observed in a spin-$2$ TLHAF \cite{Oh13}. This is consistent with $1/S$ being a measure of interaction strength---and avoided decay requiring strong interactions.

Level-continuum repulsion was also recently observed\cite{Plumb16} in the gapped spin-orbit-coupled frustrated magnet BiCu$_2$PO$_6$. This nicely fits our theoretical framework: its one-dimensional nature suggests a sharp discontinuous onset of the bare two-magnon DOS ($\gamma^2 \nu(E_{th} + \delta E) \sim 1/\sqrt{\delta E}$), preventing a smooth quasiparticle entry into the continuum. This is in contrast to the quasiparticle decay observed\cite{Stone06} in the two-dimensional PHCC. Since the latter is spin-rotation symmetric, our earlier argument implies the effective dimensional shift $D=2 \to D = 4$. Hence, $\gamma^2 \nu(E_{th} + \delta E) \sim \delta E$, consistent with the smooth entry in Fig.~\ref{fig:exp}(b).


Lastly, we consider the iconic quasiparticle dispersion of superfluid helium, Fig.~\ref{fig:exp}(c). While it was originally\linebreak

\onecolumngrid

\twocolumngrid

\noindent thought that the quasiparticle would enter the two-roton continuum \cite{Pitaevskii59}, it is now known that the dispersion instead flattens off, consistent with the discontinuous onset of the two-roton DOS \cite{Woods73,Glyde18,Suppl}. 
Here, we add the following  quantitative insights. First, the distance to the continuum is exponentially small in the bare energy (red curve). Second, the quasiparticle weight $Z$ decays to zero \emph{linearly} with this distance; the high-quality data of Refs.~\onlinecite{Donnelly1981,Glyde98,Gibbs99,Azuah13} allows us to extract this information to confirm this prediction, see inset of Fig.~\ref{fig:exp}(c). 
In fact, these two seemingly unrelated predictions are unified in our theory via the Hellmann-Feynman theorem, which yields $\mathrm d E^*/ \mathrm d E_b = Z$ (see Methods).

In conclusion, this shows that away from the universal low-energy regime, the excitations of many-body systems are not as unstructured as perhaps expected. Aside from the general message that interactions can prevent or even \emph{undo} quasiparticle decay, our model can be used to derive functional relationships between \emph{a priori} unrelated quantities,  
to extract fundamentally interesting information such as the strength of interactions from experiment---as showcased for superfluid helium. 
Our work also implies  that the existence of quasiparticle decay is not the default option, but instead places considerable constraints on  underlying physical processes.  

All of these insights taken together suggest the possibility of using interactions to control, in particular to stabilise, the behaviour of quantum matter by employing, rather than combatting, strong interactions.

\vspace{10pt}

\section*{Acknowledgements}
We thank Alexander Chernyshev, Radu Coldea, Ivan Khaymovich and Siddharth Parameswaran for discussions. RV was supported by the German Research Foundation (DFG) through the Collaborative Research Center SFB 1143 and FP acknowledges the support of the DFG Research Unit FOR 1807 through
grants no. PO 1370/2-1, TRR80, the Nanosystems Initiative Munich (NIM) by the German Excellence Initiative,  and the European Research Council (ERC) under the European Union's Horizon 2020 research and innovation program (grant agreement no. 771537). This research was conducted in part at the KITP, which is supported by NSF Grant No. NSF PHY-1748958.

\bibliography{main_arxiv.bbl}

\section*{Methods}

{\small \textbf{Exactly solvable model.} We couple a bare state $|\psi\rangle$ with bare energy $E_b$ to a continuum of states $|\varphi_\alpha\rangle$ with bare energies $E_\alpha$. I.e. $\hat H = \hat H_0 + \gamma \hat V$, where
\begin{align}
\hat H_0 &= E_b \; |\psi \rangle \langle \psi| + \int \mathrm d \alpha \;  E_\alpha |\varphi_\alpha \rangle \langle \varphi_\alpha |, \label{eq:toy1}.  \\
\hat V &= \int \mathrm d \alpha \; \left( |\psi\rangle \langle \varphi_\alpha| + |\varphi_\alpha \rangle \langle \psi|  \right)  \label{eq:toy2}.
\end{align}
The continuous label $\alpha$ satisfies $\langle \varphi_\alpha | \varphi_\beta \rangle = \delta(\alpha-\beta)$ and the density of states of the continuum is denoted as $\nu(E)$. For convenience, we define our origin to be at the onset of the continuum (i.e. in the notation of the main text, $E_{th} = 0$).

It is useful to consider the single-particle Green's function $G(E) = \langle \psi | (E-\hat H)^{-1} |\psi \rangle$. One can \emph{nonperturbatively} derive that $G(E)^{-1} = E - E_b - \gamma^2 g(E)$ where we have defined $g(E) := \int \frac{\nu(\varepsilon)}{E-\varepsilon} \mathrm d \varepsilon$. A detailed derivation can be found in the Supplemental Materials. Note that $\lim \limits_{E \to -\infty} G(E)^{-1} = -\infty$ and $\lim \limits_{E\to 0^-} G(E)^{-1} = -E_b - \gamma^2 g(0^-)$. Since $G'(E) > 0$, the existence of a (unique) pole at $E^*$ below the continuum (i.e. $E^* < 0$) is equivalent to $G(0^-)^{-1}>0$, which on its turn is equivalent to $\gamma^2 > E_b / |g(0^-)|$. If $\nu(0^+) \neq 0$ (i.e. the DOS has a discontinuous onset), then the integral defining $|g(0^-)|$ diverges, hence \emph{any} nonzero $\gamma$ will give rise to a pole below the continuum. We note that an equivalent treatment can be found in Ref.~\onlinecite{Gaveau95}.

To obtain the single-particle weight $Z = |\langle \psi|\psi^* \rangle|^2$ (where $|\psi^*\rangle$ is the wavefunction with energy $E^*<0$), consider that the weight of the delta function $\delta(E - E_b - \gamma^2 g(E))$ is given by the inverse derivative of its argument, i.e. $Z = \frac{1}{1 - \gamma^2 g'(E^*)}$. Moroever, for large $|\gamma|$, we have the relationship $E^* = \gamma^2 g(E^*)$. In particular, from this one can derive that $E^* \to -\infty$ as $|\gamma|\to \infty$. We thus have that
\begin{equation}
    \lim \limits_{|\gamma| \to \infty} Z = \lim \limits_{E \to -\infty} \left( 1 - \frac{E \; g'(E)}{g(E)} \right)^{-1}. \label{eq:Z}
\end{equation}

To evaluate this, we need the asymptotic behaviour of $g(E)$. If $\nu(E)$ has finite support, then
$g(E) \sim \frac{1}{E} \int \nu(\varepsilon) \mathrm d\varepsilon$ as $|E| \to \infty$. Plugging this into Eq.~\eqref{eq:Z}, we obtain $Z \to \sfrac{1}{2}$ as claimed in the main text.

If $\nu(E)$ is not bounded but instead decays as $\nu(E) \sim \beta/E^\alpha$ with $\alpha >0$ as $E \to + \infty$, then by the theory of Stieltjes transforms \cite{Zimering69}
\begin{equation}
g(E) \sim_{E \to - \infty} \left\{ \begin{array}{ccl}
- \tilde \beta/|E|^{\min(1,\alpha)} & &\textrm{if } 0 < \alpha \neq 1\\
\tilde \beta ( \ln |E| ) /E & & \textrm{if } \alpha = 1
\end{array} \right. \quad (\tilde \beta > 0). \label{eq:asymp_g}
\end{equation}
From these asymptotics, we obtain
\begin{equation}
\lim_{|\gamma| \to \infty} Z = \left\{ \begin{array}{lll}
\sfrac{1}{2} & & \textrm{if } \alpha\geq 1,\\
\sfrac{1}{(1+\alpha)} & &\textrm{if } 0 < \alpha < 1.
\end{array} \right. 
\end{equation}
Note that this is lower bounded by $\sfrac{1}{2}$. In particular, for $\nu(E) \propto 1/\sqrt{E}$, we obtain $Z \to \sfrac{2}{3}$ as $|\gamma| \to \infty$.

In Fig.~\ref{fig:analytical}, we plot the weight of the bare state $|\psi\rangle$ on the excited states, i.e. $\mathcal A(E) := \sum_n |\langle\psi|n\rangle|^2 \; \delta(E-E_n)$. We calculate it from the identity $\mathcal A(E) = \frac{1}{\pi} \textrm{Im} G(E-i0^+)$. A straightforward calculation (included in the Supplemental Materials) gives
\begin{equation}
\mathcal A(E) = \left\{ \begin{array}{lll} \frac{1}{\pi} \; \frac{\Gamma(E)}{(E-E_b-\gamma^2 g(E))^2 + \Gamma(E)^2} & & \textrm{if } \nu(E) \neq 0, \\
\delta(E-E_b-\gamma^2 g(E)) & & \textrm{if } \nu(E) = 0, \end{array} \right.
\label{eq:A}
\end{equation}
where $\Gamma(E) := \gamma^2 \pi \nu(E)$. Within the continuum (i.e. $\nu(E) \neq 0$), Eq.~\eqref{eq:A} can qualitatively be interpreted as a Lorentzian with an energy-dependent HWHM $\Gamma(E)$, and an energy-dependent mean $E_b + \gamma^2 g(E)$. Note that for $g(E)$ to be well-defined in the continuum, one has to interpret it as a \emph{Cauchy principal value}.

More precisely, for the left column of Fig.~\ref{fig:analytical}, we consider the DOS
\begin{equation}
\nu(E) = \left\{ \begin{array}{lll}
0 & & \textrm{if } E<0, \\
\nu_0/\sqrt{E} & & \textrm{if } E>0,
\end{array} \right. \label{eq:DOS_1D}
\end{equation}
which is what one expects for the two-particle continuum of a one-dimensional gapped model\cite{Suppl}. A straight-forward calculation gives
\begin{equation}
g(E) =  \nu_0 \int_0^\infty \frac{\mathrm d \varepsilon}{\sqrt{\varepsilon} (E -\varepsilon)} = \left\{ \begin{array}{lll}
-\frac{\pi \nu_0}{\sqrt{-E}} & & \textrm{if } E<0, \\
0 & & \textrm{if } E>0.
\end{array} \right.
\end{equation}
We set $\nu_0 = 1$. In the top panel we take $\gamma=0.2$, whereas in the second panel, $\gamma=0.7$. We consider the hypothetical scenario where the onset of the continuum is at $\omega_\textrm{min} = 2-\cos(k)$, where $k$ can physically be thought of as (total) momentum. Moreover, we take the bare level to be flat, $\omega_b = 2$. In terms of our earlier variable, where the DOS has its onset at $E=0$, we can thus say that $E_b = \omega_b - \omega_\textrm{min} = \cos(k)$.

For the right column of Fig.~\ref{fig:analytical}, we consider the DOS
\begin{equation}
\nu(E) = \left\{ \begin{array}{lll}
0 & & \textrm{if } E<0 \textrm{ or } E_m < E, \\
\nu_0 \sqrt{ E(E_m-E)} & & \textrm{if } 0 \leq E \leq E_m,
\end{array} \right. \label{eq:DOS_3D}
\end{equation}
which is what one expects for the two-particle continuum of a three-dimensional gapped model\cite{Suppl}.
This has a square-root onset at $E=0$ and a square-root termination at $E=E_m$. We obtain
\begin{equation}
g(E) = \left\{
\begin{array}{ll} \pi \nu_0 \left( E - E_m/2 \right) & \textrm{if } 0 < E < E_m,\\
\pi \nu_0 \left( E - E_m/2 - E \sqrt{ 1-E_m/E } \right) & \textrm{otherwise}.
\end{array} \right.
\end{equation}
Given our earlier results, we know that there will not \emph{always} be an isolated state below the continuum. Instead, there is a threshold value $\gamma_\textrm{th} = \sqrt{ E_b /|g(0^-)|} = \sqrt{2E_b / (\pi \nu_0 E_m)}$. If $E_b > 0$, an isolated state exists below the continuum if and only if $|\gamma|> \gamma_\textrm{th}$.

We again consider $\nu_0 = 1$, $\omega_\textrm{min} = 2-\cos(k)$ and $\omega_b = 2$, but now we also have to choose an upper threshold energy: $\omega_\textrm{max} = 5+\cos(k)$. The top panel has $\gamma=0.2$, whereas the bottom panel has $\gamma=0.5$. We note that the minimum interacting strength for which there is a state below the continuum for \emph{all} values of $k$ is $\gamma = \sqrt{\frac{2}{\pi \nu_0}} \times \sqrt{1/(2+3 \sec(k))} \big|_{k = 0} = \sqrt{\sfrac{2}{5 \pi} } \approx 0.357$.

Finally, with regards to Fig.~\ref{fig:analytical}, we mention that we also plot the real part of complex poles when they exist. We see that their location nicely agrees with where the intensity of $\mathcal A(E)$ is largest. Moreover, the data in Fig.~\ref{fig:analytical} has been convoluted with a gaussian with $\sigma = 0.025$ (in units shown). This is to give the delta-function outside the continuum a visible width.

\textbf{Ising ladder.} In Fig.~\ref{fig:ladder}(b), we plot the dynamical spin structure factor $\mathcal S^{xx}(k,\omega) = \frac{1}{2\pi} \int \langle 0 | \hat \sigma^x_{A,-k}(t) \hat \sigma^x_{A,k}(0) |0\rangle e^{i \omega t} \mathrm d t$ of the spin-$\sfrac{1}{2}$ ladder defined in the main text. This quantity is very useful, as similarly to $\mathcal A(E)$ considered in the solvable model, it tells us about weight on energy eigenstates. More precisely, $\mathcal S^{xx}(k,\omega) = \sum_n \delta(\omega-\omega_n) |\langle n |\hat \sigma^x_{A,k} |0\rangle|^2$. We calculated these dynamical spin-spin correlations by first using DMRG to obtain the ground state \cite{White92} and subsequently time-evolving $\sigma^x_{A,k}|0\rangle$ using a matrix-product-operator-based method \cite{Schollwoeck11,Zaletel15}. We found that a timestep truncation of $dt = 0.1$ and a low bond dimension of $\chi = 30$ was enough to achieve converged results. We used linear prediction \cite{White08} and multiplication by a gaussian to soften the effects of Fourier-transforming a finite-time window. This introduces an effective broadening corresponding to a convolution with a gaussian with $\sigma = 0.055$ in the units shown in Fig.~\ref{fig:ladder}.

The values of the parameters for the top panel in Fig.~\ref{fig:ladder}(b) are $g_B = 0.5$, $J_B = 1$ and $\gamma = 0.3$. If we now ramp up the coupling strength $\gamma$, however, this effectively renormalises the parameters of the Ising chain. This is because $\hat H_{\textrm{int}}$ is not \emph{purely} an interaction term: it contains an $\hat S_z$ on the Ising chain, which attempts to condense the domain walls and cause a phase transition. To prevent this, whilst ramping up $\gamma$ we also change the parameters $J_B$ and $g_B$ such that the location of the continuum (shaded region in Fig.~\ref{fig:ladder}(b)) remains \emph{roughly} unchanged. Thus, for the bottom panel, we arrive at $g_B = 0.9$, $J=3$ and $\gamma=3.4$. The location of the continuum has been determined by numerically extracting the dispersion of a single domain wall.

\textbf{Dynamics of the TLHAF.} In Fig.~\ref{fig:TLAFMnum}(a), we consider the out-of-plane dynamical spin structure factor $\mathcal S^{yy}(\bm k,\omega) = \frac{1}{2\pi} \int \langle 0 | \hat \sigma^y_{-\bm k}(t) \hat \sigma^y_{\bm k}(0) |0\rangle e^{i \omega t} \mathrm d t$ of the Hamiltonian in Eq.~\eqref{eq:TLHAF}, where we take the $120^\circ$ order to be in the $xz$-plane. This can be obtained by the methods mentioned in the case of the Ising ladder (including linear prediction), extended to the case of cylindrical geometry; for more details, see Refs.~\onlinecite{Gohlke17,Verresen18}. For the data in this work, the cylinder has a circumference $L_\textrm{circ}=6$. We checked that whilst the multimagnon continuum still had a dependence on $L_\textrm{circ}$, the single-magnon dispersion is better converged in $L_\textrm{circ}$---at least for the middle- and high-energy modes of interest. One way we checked this is by comparing energies at points which are equivalent in 2D but \emph{not} on the cylinder geometry, and finding that they agree.

Due to the absence of continuous symmetry in the ground state, the large coordination number of the lattice, and the fact that the isotropic point has \emph{three} Goldstone modes, it is numerically challenging to time-evolve this highly-entangled state. For this reason we are limited in the bond dimensions that we can reach: $\chi=450$ for long-time dynamics necessary for resolving high-energy modes, and $\chi=800$ for short-time dynamics for low-energy modes (see discussion below).

The numerical parameters for Fig.~\ref{fig:TLAFMnum}(a) correspond to a timestep truncation $dt = 0.05J$, a bond dimension $\chi = 450$, and an effective gaussian broadening with $\sigma = 0.077J$. The dotted line in Fig.~\ref{fig:TLAFMnum}(a) is the sum $\varepsilon_{\bm q} + \varepsilon_{\textrm K}$, where $\bm q$ is along the orange line in the inset. Here $\varepsilon_{\bm q}$ was obtained by tracing the peak of the spectral function along that slice; $\varepsilon_{\textrm{K}}$ is a low-energy feature which could not be resolved with the bond dimension $\chi=450$. Instead, we went up to $\chi =800$, which limited the time-window we could obtain, leading to a larger effective broadening. However, since the low-energy mode is well-separated from other (relevant) modes, one can still reliably extract the energy from a broad response. From a scaling in bond dimension, we then obtained $\varepsilon_{\textrm K} \approx 0.3J$ for the value $\delta = 0.05$. This extrapolation is represented visually in the Supplemental Materials\cite{Suppl}. This is markedly lower than the LSWT prediction, $\varepsilon_{\textrm K}^\textrm{LSWT} \approx 0.41J $.

The magnon dispersion in Fig.~\ref{fig:TLAFMnum}(c) was obtained by tracing the low-energy peak of the spectral function---having verified that the magnon branch was resolved enough for this to be sensible. At low energies, this was supplemented by the aforementioned approach where we could go up to $\chi=800$.
Due to the cylindrical geometry on which our method is based, the dispersion we obtain is continuous along one direction, and discrete along the other. We then superimposed the momentum cuts along three different orientations and subsequently interpolated this to the full two-dimensional Brillouin zone\cite{Suppl}. The fact that where these cuts intersected, they agreed, is a confirmation that the circumference $L_\textrm{circ}=6$ is large enough for the single-magnon dispersion to resemble the true two-dimensional result. As a sanity check for our interpolation method, we have verified that it gives the correct result when applied to the LSWT dispersion, as shown in the Supplemental Materials\cite{Suppl}.

\textbf{Experimental data for the TLHAF.} In the inelastic neutron scattering data for Ba$_3$CoSb$_2$O$_9$ in Fig.~\ref{fig:exp}(a), the momentum-cut is along K--K'. In the inset of Fig.~\ref{fig:TLAFMnum}(a), K' is shown as a corner point of the (first) BZ. However, in the experiment\cite{Ito17}, K' was taken in the second BZ (which differs from the other choice by a reciprocal lattice vector). This difference has no bearing on the bands one picks up, so for our purposes this distinction is irrelevant. It does, however, affect the precise value of the intensity. This explains why Fig.~\ref{fig:exp}(a) is not left-right symmetric.

\textbf{Subtleties near and at the isotropic point of the TLHAF.} The decay process $\bm k \to \textrm{K} + (\bm k - \textrm K )$ accounts for the \emph{complete} decay region (as predicted by LSWT) only at the isotropic point (i.e $\delta = 0$). For $\delta \neq 0$, this process represents the core of the decay region, which is then slightly extended by considering $\bm k \to \bm q + (\bm k - \bm q )$ with $\bm q \approx \textrm K$. One consequence is that the minimum predicted by the principle of avoided decay is only precisely at Y$_1$ at the isotropic point. Indeed, in Fig.~\ref{fig:TLAFMnum}(c) one can see that the minimum (for $\delta =0.05$) has been slightly shifted inward, albeit not very substantially so.

Interestingly, at the isotropic point $\delta=0$, absence of decay is \emph{equivalent} to the magnon dispersion $\varepsilon_{\bm k}$ being periodic with respect to the magnetic BZ---which is three times smaller than the original BZ. (This can be derived from the fact that $\varepsilon_\textrm{K} = 0$ for $\delta = 0$.) This powerful criterion might help to figure out the extent of (avoided) decay at the isotropic point, be it using numerical or experimental methods.

\textbf{Relationship between $\bm{E_b}$, $\bm{E^*}$ and $\bm Z$.} In the main text, we alluded to the \emph{general} relationship $\mathrm d E^* / \mathrm d E_b = Z$. This is a general property of our model. To prove this, first rewrite 
\begin{equation}
    \frac{\mathrm d E^*}{\mathrm d E_b} = \frac{\mathrm d}{\mathrm d E_b} \langle \psi^* | \hat H | \psi^* \rangle  =  \langle \psi^* | \frac{\mathrm d \hat H}{\mathrm d E_b}| \psi^* \rangle, \label{eq:Hellmann}
\end{equation}
where we used the Hellmann-Feynman theorem to move the derivative inside. The proof is finished by noting that Eq.~\eqref{eq:toy1} implies $\frac{\mathrm d \hat H}{\mathrm d E_b} = |\psi \rangle \langle \psi|$.

\textbf{Predictions for helium.} Lastly, we make a few comments relevant to the case of superfluid helium. As shown in the Supplemental Materials, the two-roton continuum has a jump discontinuity\cite{Suppl}. Hence, let us consider the case where $\nu(E)$ has a discontinuous onset $\nu_0$. Then a straight-forward computation shows that $g(E) \sim \nu_0 \ln (-E) + \textrm{const}$, for $E$ small and negative. Hence, remembering the condition we derived above ($E^* = E_b + \gamma^2 g(E^*)$), we see that as $E^*\to 0^-$, we have the functional relationship $\nu_0 \ln(-E^*) = E_b + \textrm{const}$, i.e. $E^* \propto \exp\left( - E_b / \nu_0 \right)$. Using the fact\cite{Suppl} that $\nu_0 \sim 1/k$, we obtain the formula mentioned in the main text. Using the general relationship $\mathrm d E^* / \mathrm d E_b = Z$, we also directly obtain the other prediction. In particular, this means that the values of $a$ and $b$ (the parameters mentioned in the main text) should be equal. However, it does not make sense to expect this for the experimental data, as the weight $Z$ extracted in that setting is usually only defined up to a global (momentum-independent) multiplicative factor.

\textbf{Experimental data for helium.} With regard to the experimental data for helium, the quasiparticle dispersion relation was straightforwardly extracted from Refs.~\onlinecite{Donnelly1981,Glyde98}. The weight, however, is more subtle: Refs.~\onlinecite{Gibbs99,Azuah13} showed the data as a function of momentum, which we extracted and interpolated. We then evaluated this interpolated function at the same momenta for which Refs.~\onlinecite{Donnelly1981,Glyde98} quoted values for the energy. This allowed us to plot $Z$ as a function of energy in the inset of Fig.~\ref{fig:exp}(c).
}

\vspace{30pt}

\pagebreak

\widetext
\ifx\targetformat\undefined
\begin{center}
	\textbf{\large Supplemental Materials: ``Strong quantum interactions prevent quasiparticle decay''}
\end{center}

\setcounter{equation}{0}
\setcounter{figure}{0}
\setcounter{table}{0}
\makeatletter
\renewcommand{\theequation}{S\arabic{equation}}
\renewcommand{\thefigure}{S\arabic{figure}}
\renewcommand{\bibnumfmt}[1]{[S#1]}
\renewcommand{\citenumfont}[1]{S#1}

\section{The interacting single-particle Green's function and spectral function \label{app:Green}}

We will first calculate $G(E) = \langle \psi | (E-\hat H)^{-1}|\psi\rangle$.
If we think of $E-\hat H$ as a matrix (with indices labeled by $|\psi\rangle$ and $|\varphi_\alpha\rangle$), then $G(E)$ is the top left element of its inverse. This is easily calculated. Schematically, first write
\begin{equation}
E-\hat H = \left( \begin{array}{cc} A & B\\
C & D
\end{array} \right) \textrm{ with } A = E-E_b, \quad D_{\alpha\beta} = (E-E_\alpha) \; \delta(\alpha-\beta)  \quad \textrm{and } B_\alpha = C_\alpha = - \gamma.
\end{equation}
Since $D$ is diagonal, one can apply the well-known result that \begin{equation}
\det(E-\hat H) = \det(A - B D^{-1}C) \det D = \left( E - E_b - \gamma^2 \int \mathrm d \alpha \frac{1}{E-E_\alpha} \right) \det D.
\end{equation}
Finally, by Cramer's rule, we can express
\begin{equation}
G(E) = \langle \psi | (E-\hat H)^{-1} |\psi \rangle = \frac{\det D}{\det (E-\hat H)} = \frac{1}{ E - E_b - \gamma^2 g(E) },
\end{equation}
where we have introduced the Cauchy principal value $g(E) := \int_0^\infty \frac{\nu(\varepsilon)}{E-\varepsilon} \mathrm d \varepsilon$, as in the main text.

Note that 
\begin{equation}
g(E-i \eta ) = \int_0^\infty \nu(\varepsilon) \frac{E-\varepsilon +i \eta}{(E-\varepsilon)^2 + \eta^2} \mathrm d \varepsilon, \textrm{ hence } \frac{1}{\pi} \textrm{Im} \; g(E-i0^+)= \int_0^\infty \nu(\varepsilon) \; \delta(E-\varepsilon) = \nu(E).
\end{equation}
In other words, $g(E-i0^+) = g(E) + i \pi \nu(E)$.

We can now calculate $\mathcal A(E) = \frac{1}{\pi} \textrm{Im} G(E-i0^+)$ as follows,
\begin{align}
\mathcal A(E) &= \frac{1}{\pi} \textrm{Im} \left( \frac{1}{E-E_b-\gamma^2 g(E-i0^+) -i0^+} \right) \\
&= \frac{1}{\pi} \textrm{Im} \left( \frac{1}{E-E_b-\gamma^2 g(E) - i \pi \gamma^2 \nu(E)-i0^+} \right) \\
&= \left\{ \begin{array}{c c c} \frac{\gamma^2 \nu(E)}{(E-E_b-\gamma^2 g(E))^2 + (\pi \gamma^2 \nu(E))^2} & & \textrm{if } \nu(E) \neq 0, \\
\delta(E-E_b-\gamma^2 g(E)) & & \textrm{if } \nu(E) = 0. \end{array} \right.
\end{align}

\section{two-particle DOS}

\subsection{Quadratic dispersion}
Here we calculate the two-particle DOS in various dimensions $D$ for the single-particle dispersion
\begin{equation}
\varepsilon_{\bm k} = \Delta + \frac{|\bm k |^2}{2m}.
\end{equation}

\subsubsection{\texorpdfstring{$D = 1$}{D=1}}

\begin{align}
    \rho_2^{(1D)}(k,\varepsilon) &\propto \int \delta(\varepsilon_q + \varepsilon_{k-q} - \varepsilon ) \; \mathrm d q
    = \int \delta \left( 2\Delta - \varepsilon + \frac{q^2}{2m} + \frac{(k-q)^2}{2m} \right) \; \mathrm d q \\
    & = \int \delta \left( 2\Delta - \varepsilon + \frac{1}{2m}\left[ 2 \left( q-\frac{k}{2} \right)^2 + \frac{k^2}{2} \right] \right) \; \mathrm d q \\
    &\propto \left. \frac{1}{\frac{1}{m} \times \left|q- \frac{k}{2} \right|} \right|_{|q-k/2| = \sqrt{m(\varepsilon-2 \Delta) - k^2/4} } = \frac{\sqrt{m}}{\sqrt{\varepsilon-2\Delta - k^2/4m}}.
\end{align}

\subsubsection{\texorpdfstring{$D = 2$}{D=2}}

\begin{align}
\rho_2^{(2D)}(\bm k, \varepsilon)
&\propto \int \delta(\varepsilon_{\bm q} + \varepsilon_{ \bm k - \bm q} - \varepsilon ) \; \mathrm d^2 \bm q
 = \int \delta \left( 2\Delta - \varepsilon + \frac{q_x^2}{2m}+\frac{q_y^2}{2m} + \frac{(k_x-q_x)^2}{2m}+\frac{(k_y-q_y)^2}{2m} \right) \; \mathrm d q_x\mathrm d q_y \\
&= \int \rho_2^{(1D)}\left( k_x, \; \varepsilon - \frac{q_y^2}{2m} - \frac{(k_y-q_y)^2}{2m} \right) \; \mathrm d q_y
= \int \frac{\sqrt{m}}{\sqrt{\varepsilon - 2 \Delta - q_y^2/2m - (k_y-q_y)^2/2m - k_x^2/ 4m}} \; \mathrm d q_y \\
&= \int \frac{m}{\sqrt{\left( \varepsilon - 2 \Delta - (k_x^2 + k_y^2)/4m \right) - (q_y - k_y/2)^2}}\; \mathrm d q_y 
\propto \left\{
\begin{array}{lll}
m & & \textrm{if } \varepsilon > 2\Delta + \frac{k_x^2 + k_y^2}{4m}, \\
0 & & \textrm{otherwise.}
\end{array}
\right.
\end{align}

\subsubsection{\texorpdfstring{$D \geq 3$}{D>=3}}

\begin{equation}
\rho_2^{(3D)}(\bm k, \varepsilon)
\propto \int \delta(\varepsilon_{\bm q} + \varepsilon_{ \bm k - \bm q} - \varepsilon ) \; \mathrm d^D \bm q
\propto \int \left. \left( \frac{1}{\left|\partial_\theta \varepsilon_{\bm k - \bm q} \right|} \;\sin \theta \right) \right|_{\varepsilon_{\bm q} + \varepsilon_{ \bm k - \bm q} = \varepsilon} \; q^{D-1} \mathrm d q \label{eq:rho2spherical}
\end{equation}

Since $|\bm k - \bm q |^2 = k^2 + q^2 - 2kq \cos \theta$, we have that $\partial_\theta \varepsilon_{\bm k - \bm q} = \frac{kq}{m} \sin \theta$. Hence $\rho_2(\bm k, \varepsilon) \propto \frac{m}{k} \int q^{D-2} \; \mathrm dq$. To determine the range of integration, it is useful to first define $\delta$ through $\varepsilon = 2\Delta + \frac{k^2}{4m} + \frac{\delta}{m}$. From the condition that $\varepsilon = \varepsilon_{\bm q} + \varepsilon_{ \bm k - \bm q}$ and that $|\cos \theta| \leq 1$, we obtain the condition on $q$, i.e. $|\sqrt{\delta} - k/2| \leq q \leq \sqrt{\delta} + k/2$. Note that this only makes sense if $\delta \geq 0$, i.e. it is the correct variable to use to describe the onset of the DOS. Plugging this in and using that $\delta$ is small, we obtain
\begin{align}
\rho_2(\bm k, \varepsilon)
&\propto \frac{m}{k} \; \left. q^{D-1} \right|^{ \sqrt{\delta} + k/2 }_{ | \sqrt{\delta} - k/2 | }
\propto \frac{m}{k} \; k^{D-1} \; \left[ \left(1+ \frac{\sqrt{\delta}}{k} \right)^{D-1} - \left( 1- \frac{\sqrt{\delta}}{k}\right)^{D-1} \right] \\
&\approx \frac{m}{k} \; k^{D-1} \; \left[ \left(1+ (D-1)\frac{\sqrt{\delta}}{k} \right) - \left( 1- (D-1)\frac{\sqrt{\delta}}{k}\right) \right] 
\approx m k^{D-3} \sqrt{\delta} = m^{3/2} k^{D-3} \sqrt{ \varepsilon-2\Delta - \frac{k^2}{4m} }.
\end{align}
Hence, for $D \geq 3$, the onset always has a square-root onset. \emph{However}, the above derivation is only for a narrow window near the onset, and this window vanishes as one approaches $\bm k \to 0$. Indeed, one can straightforwardly calculate that 
\begin{equation}
\rho_2(\bm k = 0, \varepsilon) \propto \int \delta \left( 2\Delta- \varepsilon + \frac{q^2}{m} \right) \; q^{D-1} \mathrm d q \propto \left. m q^{D-2} \right|_{q^2 = m(\varepsilon-2\Delta)} = m^{\sfrac{D}{2}} \; (\varepsilon-2\Delta)^{\sfrac{D}{2}-1}.
\end{equation}
This is physically the behaviour that will dominate near $\bm k \approx 0$.

\subsection{Roton minima}
We now consider the dispersion relevant to the roton minimum appearing in, for example, superfluid helium,
\begin{equation}
\varepsilon_{\bm k} = \Delta + \frac{1}{2m}(|\bm k | - K)^2.
\end{equation}

Here we restrict ourselves to $D\geq 3$, where we can use Eq.~\eqref{eq:rho2spherical}. Since $|\bm k - \bm q | = \sqrt{k^2 + q^2 - 2kq \cos \theta}$, we have that \begin{equation}
\partial_{\theta} \varepsilon_{\bm k - \bm q} = \sfrac{1}{m} \times (|\bm k - \bm q|-K) \times \frac{1}{2|\bm k - \bm q|} \times 2 kq \sin \theta .
\end{equation}

We are interested in $0 < k < 2K$, where the threshold is near $\varepsilon \approx 2\Delta$, which forces the decay products to be very close to the roton minimum, i.e $q \approx K$ and $|\bm k - \bm q| \approx K$. Hence, near the threshold we have
\begin{align}
\rho_2(\bm k, \varepsilon)
&\propto m \int \left. \left( \frac{|\bm k - \bm q|}{\left| |\bm k -\bm q|-K \right|} \right) \right|_{\varepsilon_{\bm q} + \varepsilon_{ \bm k - \bm q} = \varepsilon} \; \frac{q^{D-2}}{k} \; \mathrm dq 
\approx \frac{m K^{D-1}}{k} \int \left. \left( \frac{1}{ \sqrt{ 2m \left( \varepsilon_{\bm k - \bm q} - \Delta \right) } }  \right) \right|_{\varepsilon_{\bm q} + \varepsilon_{ \bm k - \bm q} = \varepsilon} \; \mathrm dq \\
& = \frac{m K^{D-1}}{k} \int_{K - \sqrt{2m(\varepsilon - 2 \Delta)}}^{K + \sqrt{2m (\varepsilon - 2 \Delta)}}  \frac{1}{\sqrt{2m (\varepsilon - 2 \Delta) - (q-K)^2 }} \mathrm d q \\
& = \frac{m K^{D-1}}{k} \int_{- 1}^{1}  \frac{1}{\sqrt{1 -x^2}} \mathrm d x \quad \left(\textrm{where } x := \frac{q-K}{\sqrt{2m (\varepsilon - 2 \Delta)}}\right) \\
&\propto \left\{ \begin{array}{ccc}
0 & & \textrm{ if } \delta<0 \\
\frac{m K^{D-1}}{k} & & \textrm{ if } \delta>0
\end{array} \right. \qquad \textrm{ where } \varepsilon \approx 2\Delta + \delta.
\end{align}
We repeat that the above derivation is for $0< k < 2K$. We conclude that in this regime, there is a jump discontinuity at the onset.

\section{DMRG analysis of 2D TLHAF}


\begin{figure}[h]
    \includegraphics{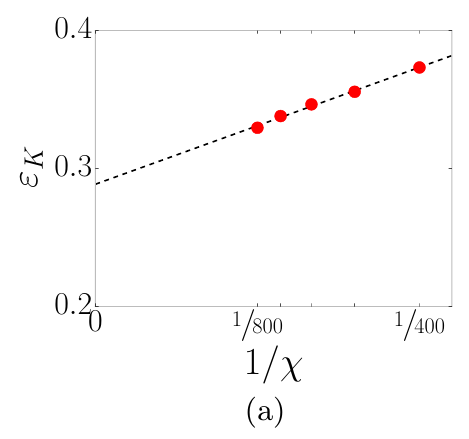}
    \includegraphics{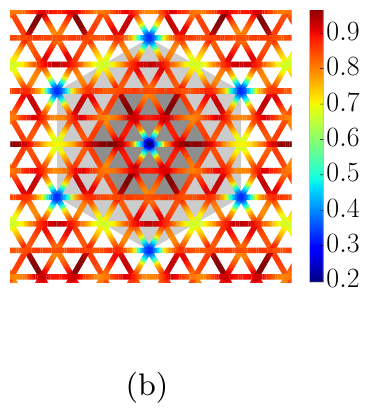}
    \includegraphics{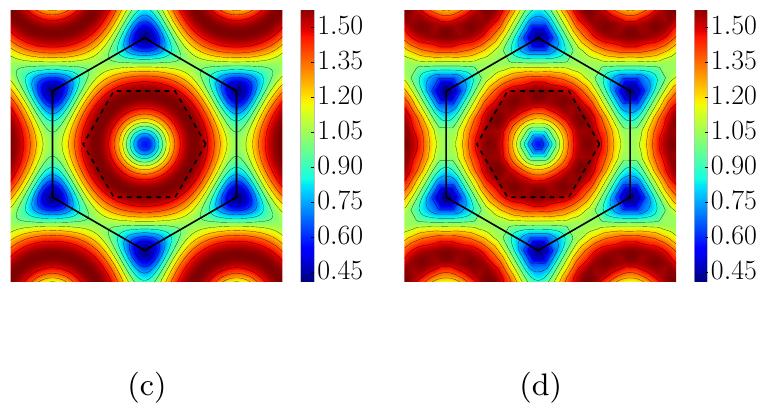}
    \caption{(a) By scaling in bond dimension ($\chi = 400,500,600,700,800$), we find that $\varepsilon_K\approx 0.3 J$ for $L_\textrm{circ}=6$ and $\delta = 0.05$. (b) The dispersion for $\delta=0.05$ that we obtained numerically (in units of $J$) before interpolating to 2D. (c) The LSWT prediction for $\delta=0.05$ (in units of $J$). (d) The result of first restricting the aforementioned LSWT prediction onto the grid shown in (b) and then using our 2D interpolation  method; the fact that this closely agrees with (c) is an indication that our interpolation method is reliable. \label{fig:suppl}}
\end{figure}

Fig.~\ref{fig:suppl}(a) shows how by scaling in bond dimension, we can get an estimate $\varepsilon_K\approx 0.3J$ for $\delta=0.05$.  

Fig.~\ref{fig:suppl}(b) shows the data that we can numerically obtain for the magnon dispersion for a cylinder with circumference $L_\textrm{circ}=6$, where we have rotated and superimposed the data along three different directions. This is for $\delta=0.05$. The fact that the values roughly agree when they spatially overlap indicates that the finite-circumference effects are not too strong. We interpolated this data to the two-dimensional BZ to generate Fig.~\ref{fig:TLAFMnum}(c) in the main text.

To test our interpolation method, we can take the LSWT prediction (Fig.~\ref{fig:suppl}(c)) as a test case: restricting this to the same grid as is shown in Fig.~\ref{fig:suppl}(b) and then using our 2D interpolation method, we produce Fig.~\ref{fig:suppl}(d). We see that this closely agrees with the original dispersion in Fig.~\ref{fig:suppl}(c).
\end{document}